\documentclass[11pt]{article}
\usepackage{euler}
\usepackage{ams}
\usepackage{xypic}
\newtheorem{tm}{Theorem}

\newtheorem{cha}{Challenge}

\input xy
\xyoption{all}

\begin{document}
\begin{center}{\Large{\bf A New Key-Agreement-Protocol}}\\
\vspace{.5cm}
{Bj\"orn Grohmann}\\\vspace{.2cm}
{\small Universit\"at Karlsruhe, Fakult\"at f\"ur Informatik,\\76128 Karlsruhe, Germany}\\
{\small {\tt nn@mhorg.de}}
\end{center}
\vspace{.5cm}
\begin{abstract}
A new 4-pass Key-Agreement-Protocol is presented. The security of the protocol mainly relies on the existence of a (polynomial-time computable) One-Way-Function and the supposed computational hardness of solving a specific system of equations.
\end{abstract}
\section{Introduction}
At the end of a Key-Agreement-Protocol two parties, say Alice and Bob, share a common bit string $s$. During the protocol they are allowed to exchange a fixed number of messages $m_i$, $i=1,\dots, r$, over a public channel. The protocol is called secure, if no algorithm exist that computes the string $s$ from the $m_i$'s in a polynomial number of steps. Whether secure Key-Agreement-Protocols exist is still an open issue, although quite a few have been proposed -- maybe the most popular being the Diffie-Hellman-Protocol \cite{dihe1976}, where the security is linked to the task of computing the element $\gamma^{ab}$ of a given cyclic group from  the elements $\gamma^a$ and $\gamma^b$.\\\\
In this article, we present a new Key-Agreement-Protocol that uses four rounds of message exchange. Its security mainly relies on the existence of a (polynomial-time computable) One-Way-Function and the supposed computational hardness of solving a specific system of equations.
\section{The Protocol}
{\bf Public data:} Suppose Alice and Bob want to exchange a secret key. They start by agreeing on a positive integer $n$ and a prime $p$ of size $\sim 2^{\sqrt{n\log n}}$. They further agree on a random matrix $C:=\left(c_{i,j}\right)_{i,j}\in \F_p^{n\times n}$, with
$i,j\in \{1,\dots,n\}$, and an injective (polynomial-time computable) One-Way-Function $h:\F_p\longrightarrow \{0,1\}^m$, where $\F_p$ denotes the finite field with $p$ elements.\\\\
{\bf Private data:} Next, Alice (resp. Bob) chooses a random element $\alpha \in \F_p$ 
(resp. $\beta$), $n$ random bits $t_1,\dots, t_n$ (resp. $s_1,\dots,s_n$) 
and a random permutation $\sigma$ on the set $\{1,\dots,n \}$ (resp. $\varrho$), all of which she (resp. he) keeps secret.\\\\
The computations that follow are all taking place in the finite field $\F_p$.\\\\
{\bf First round:} Alice computes for $j=1,\dots,n$:
\begin{equation}\label{eqtsa}
\mu_j:=\sum_{i=1}^{n}t_i c_{i,j}+\sigma(j)\alpha
\end{equation}
and sends $(\mu_j)_j$ to Bob.\\\\
{\bf Second round:} Bob computes for $i=1,\dots,n$:
\begin{equation}
\nu_i:=\sum_{j=1}^{n}s_j c_{i,j}+\varrho(i)\beta \,\,\mbox{ and }\,\, \tau_A:=\sum_{j=1}^{n}s_j \mu_j
\end{equation}
and sends $((\nu_i)_i,\tau_A)$ to Alice.\\\\
{\bf Third round:} Alice computes for $k=1,\dots,\frac{n(n-1)}{2}$:
\begin{equation}
h(\tau_A-k\alpha) \,\,\mbox{ and }\,\, \tau_B:=\sum_{i=1}^{n}t_i \nu_i
\end{equation}
and sends $((h(\tau_A-k\alpha))_k,\tau_B)$ to Bob.\\\\
{\bf Final round:} Bob computes for $l=1,\dots,\frac{n(n-1)}{2}$ the list $(h(\tau_B-l\beta))_l$ until he finds $k_0$ and $l_0$, such that
\begin{equation}
h(\tau_A-k_0\alpha) = h(\tau_B-l_0\beta)
\end{equation}
and sends $k_0$ to Alice.\\\\
Alice and Bob now share a common element $g:=\tau_A-k_0\alpha = \tau_B-l_0\beta$.
\section{Analysis}
We start by showing the correctness of the protcol and calculate the computational cost:
\begin{tm}
After the final step both parties share a common element $g$. The number of computational steps on both sides equals ${\bf O}(n^2\cdot \mbox{\rm cost of evaluation of }h)$.
\end{tm}
{\bf Proof.} The correctness of the protocol follows from the easy observation that
\begin{equation}
\tau_A=\sum_{i,j=1}^{n}t_i s_j c_{i,j} + \alpha \sum_{j=1}^{n}s_j \sigma(j) = g^{\prime}+\alpha k^{\prime},
\end{equation}
and respectively
\begin{equation}
\tau_B=\sum_{i,j=1}^{n}t_i s_j c_{i,j} + \beta \sum_{i=1}^{n}t_i \varrho(i) = g^{\prime}+\beta l^{\prime},
\end{equation}
and the fact that $1 \le k^{\prime},l^{\prime} \le n(n-1)/2$, which means that at least one pair of integers $(k_0,l_0)$ within the given range exists, such that $g:=\tau_A-k_0\alpha = \tau_B-l_0\beta$. The number of computational steps is also clear, since Bob can sort the list $(h(\tau_A-k\alpha))_k$ in ${\bf O}(n^2 \log n)$ steps, while the evaluation of the injective function $h$ requires $\mbox{\boldmath $\Omega$}(\log p)$ operations. 
\hfill$\Box$\\\\
The above protocol gives rise to the following
\begin{cha}\label{con1}
Given $n$, $p$, $h$, $C$, $(\nu_i)_i$, $(\mu_j)_j$, $\tau_A$, $\tau_B$, $(h(\tau_A-k\alpha))_k$ and $k_0$, compute an element $g$, such that $h(g)=h(\tau_A-k_0\alpha)$.
\end{cha}
We (i.e. the author of this article) are not aware of any lower bound for the number of steps it takes to compute the element $g$ from Challenge \ref{con1}.\\\\
In what follows, we will present an algorithm that conjecturally requires $\mbox{\boldmath $\Omega$}(2^{\varepsilon\sqrt{n\log n}})$ operations, for some constant $\varepsilon>0$.\\\\
We will try to compute the secrect bits $t_1,\dots,t_n$ of Alice. As is easily seen, the knowledge of these bits will lead in a polynomial number of steps to the secret key. At the beginning there is only one equation for these bits, that is
\begin{equation}\label{eqxn}
x_1 \nu_1 + \dots + x_n \nu_n = \tau_B.
\end{equation}
Now, heuristically speaking, while there are $2^n$ ways to select the values of the $x_i$'s but only $p\sim 2^{\sqrt{n\log n}}$ possible values for $\tau_B$, there are approximately $2^{n-\log p}\sim 2^{n(1-\sqrt{log n/n})}$ solutions to equation (\ref{eqxn}) (in the language of Knapsack-Cryptography, we could speak of an ultra-high density Knapsack, since the density of this Knapsack tends to infinity \cite{ngst2005}).\\\\
The other equations from (\ref{eqtsa}) involving the $t_i$'s can not be used immediately, since the permutation $\sigma$ and the element $\alpha$ are both secret, but we can try to get rid of $\alpha$
by guessing $r$ values of the permutation $\sigma$, say $\sigma^{\prime}(1),\dots,\sigma^{\prime}(r)$, which gives us $r-1$ additional equations:
\begin{eqnarray*}
\sum x_i (\sigma^{\prime}(2)c_{i,1}-\sigma^{\prime}(1)c_{i,2}) & = & \sigma^{\prime}(2)\mu_1-\sigma^{\prime}(1)\mu_2\\
\sum x_i (\sigma^{\prime}(3)c_{i,1}-\sigma^{\prime}(1)c_{i,3}) & = & \sigma^{\prime}(3)\mu_1-\sigma^{\prime}(1)\mu_3\\
& \vdots & \\
\sum x_i (\sigma^{\prime}(r)c_{i,1}-\sigma^{\prime}(1)c_{i,r}) & = & \sigma^{\prime}(r)\mu_1-\sigma^{\prime}(1)\mu_r.\\
\end{eqnarray*}
Again, by the same heuristic argument, the system of these equations together with equation (\ref{eqxn}) has approximately $2^{n-r\log p} \sim 2^{n(1-r\sqrt{log n/n})}$ solutions, which means that we can not even be sure whether our guess was right, unless
$n-r\log p\sim \log^{\kappa} n$, for some constant $\kappa$.\\\\
To summarize the discussion, the probability of guessing enough equations to compute the $t_i$ (where we did not even talk about the computational cost of really solving these equations) is about $n^{-\varepsilon n/\log p}\sim 2^{-\varepsilon\sqrt{n\log n}}$, for some constant $\varepsilon>0$, which is, at least from a theoretical point of view not too far away from the probability of guessing the secret $\alpha$ (resp. the secret key $g$) directly.\\\\
It is almost superfluous to say that these heuristic considerations do not prove anything about the security of the stated protocol. Nevertheless, in the author's opinion, Challenge \ref{con1} seems worth further investigation.

\end{document}